\documentclass{aastex}

\usepackage{emulateapj5}

\shorttitle{Atomic gas in FRB hosts: interactions}
\shortauthors{Micha{\l}owski}
\newcommand{\myemail}{mj.michalowski@gmail.com}
\newcommand{\urltt}[1]{\url{\texttt{#1}}}

\usepackage{color}

\newcommand{\msun}{\mbox{$M_\odot$}}
\newcommand{\msunyr}{\mbox{\msun\,yr$^{-1}$}}
\newcommand{\inst}[1]{\altaffilmark{#1}}
\usepackage{comment}

\newcommand{\mstar}{\mbox{$M_*$}}
\newcommand{\mhi}{\mbox{$M_{\rm HI}$}}

\newcommand{\hi}{\sc Hi}
\newcommand{\kms}{\mbox{km\,s$^{-1}$}}

\newcommand{\frb}{FRB\,181030A}
\newcommand{\ngc}{NGC\,3252}

\newcommand{\frbm}{FRB\,200120E}

\newcommand{\frbmw}{FRB\,200428}

\begin{document}

\title{
Asymmetric {\hi} 21\,cm lines of fast radio burst hosts: connection with galaxy interaction
}

\author{Micha{\l}~J.~Micha{\l}owski\inst{\ref{inst:uam}}
	}

\altaffiltext{1}
{Astronomical Observatory Institute, Faculty of Physics, Adam Mickiewicz University, ul.~S{\l}oneczna 36, 60-286 Pozna{\'n}, Poland \myemail  \label{inst:uam}}


\begin{abstract}
Fast radio bursts (FRBs) are enigmatic transients with very short radio emission. Their nature is still widely debated. I provide the first analysis of atomic gas properties of a small sample of FRB hosts to constrain their nature. {\hi} observations exist for {\ngc}, the host of {\frb}, M81, the host of {\frbm}, and the Milky Way, the host of {\frbmw}.
I report three observables: {\it i)} all three FRB hosts are interacting galaxies; {\it ii)} the {\hi} spectra of both FRB hosts with such data available are highly asymmetric, several standard deviations above the general population of galaxies; {\it iii)} two FRB hosts have normal atomic gas properties and one is strongly deficient in atomic gas. This indicates that nearby and repeating FRBs are connected with a recent enhancement of star formation due to interaction. This supports fast FRB channels, for example a massive star with a short delay time so that interaction signatures giving rise to the birth of the progenitor are still visible.
Long gamma-ray burst (GRB) and broad-lined type Ic supernova (SN) hosts exhibit much more symmetric spectra, even though they were  claimed to experience gas inflow from the intergalactic medium. The difference can be explained by the interactions experienced by FRB hosts being more disruptive than these gas inflows, or by the mass effect, with GRB/SN hosts at lower masses having less organized gas motions, so with {\hi} lines closer to a symmetrical Gaussian.
This also suggests that the emission mechanisms of FRBs and GRBs are different.
\end{abstract}

\keywords{
Radio bursts (1339);
Radio transient sources (2008);
Gamma-ray bursts (629);
Interstellar atomic gas (833);
HI line emission (690);
Interacting galaxies (802);
Galaxy groups (597);
Spiral galaxies (1560)
}

\section{Introduction}

Fast radio bursts (FRBs) are enigmatic transients with very short radio emission \citep{lorimer07,petroff19}. 
Their nature is still unknown and is widely debated. It is only certain that most of them are of extragalactic origin, as revealed by  their high dispersion measure values.
One FRB was shown to be associated with a Galactic magnetar suggesting that at least a fraction of FRBs is related to neutron stars \citep{bochenek20}.
There are many possible theories for the mechanism of the FRB emission \citep{platts19}. The most promising ones, related to compact objects, can be divided into two categories depending on the time elapsed since the birth of the progenitor. Slow channels involve the evolution of the order of a billion years before the burst, for example mergers of compact objects (black holes, neutron stars, or white dwarfs) or interactions of compact objects in close binary systems. Alternatively, fast channels give rise to the burst in a few tens to a hundred of million years, for example supernova (SN) explosions in binary systems, SN remnants (pulsars and magnetars), or collapses of compact objects.

The importance of FRBs is their potential use to probe extreme physical processes in the Universe and the otherwise invisible intervening medium \citep{mcquinn14,petroff19,macquart20}.  In order to do the former, one needs first to understand the mechanism of their emission. 

Some clues to the nature of these unusual objects can be obtained from the properties of gas in their environment. In this way, strong atomic concentrations found near the positions of long-duration gamma-ray bursts (GRBs) and broad-lined type Ic (Ic-BL) SNe were used to advocate that their progenitors were born in a star formation episode triggered by a recent inflow of gas from the intergalactic medium or galaxy interaction \citep{michalowski15hi,michalowski18,michalowski20,arabsalmani15b,arabsalmani19}. On the other hand, the lack of such concentrations for the enigmatic transient AT\,2018cow \citep{michalowski19,roychowdhury19} and the SN-less GRB\,111005A \citep{lesniewska21}, suggests that their mechanism may be different from the explosion of a very massive star.


Properties of FRB hosts derived from their stellar emission have been used to learn about FRBs with mixed results \citep{nicholl17,bhandari20,heintz20,li20b,marcote20,fong21,kirsten21,mannings21,piro21,tendulkar21}.
None of these works discussed cold gas properties, so the aim of this Letter is to provide the first analysis of atomic gas properties of FRB hosts to provide constraints on their nature.
The sample size is limited, so to obtain fully statistically significant conclusions more nearby FRBs need to be identified or more distant hosts need to be observed with future radio telescopes.


\begin{figure*}
\begin{tabular}{cc}
\includegraphics[width=0.5\textwidth]{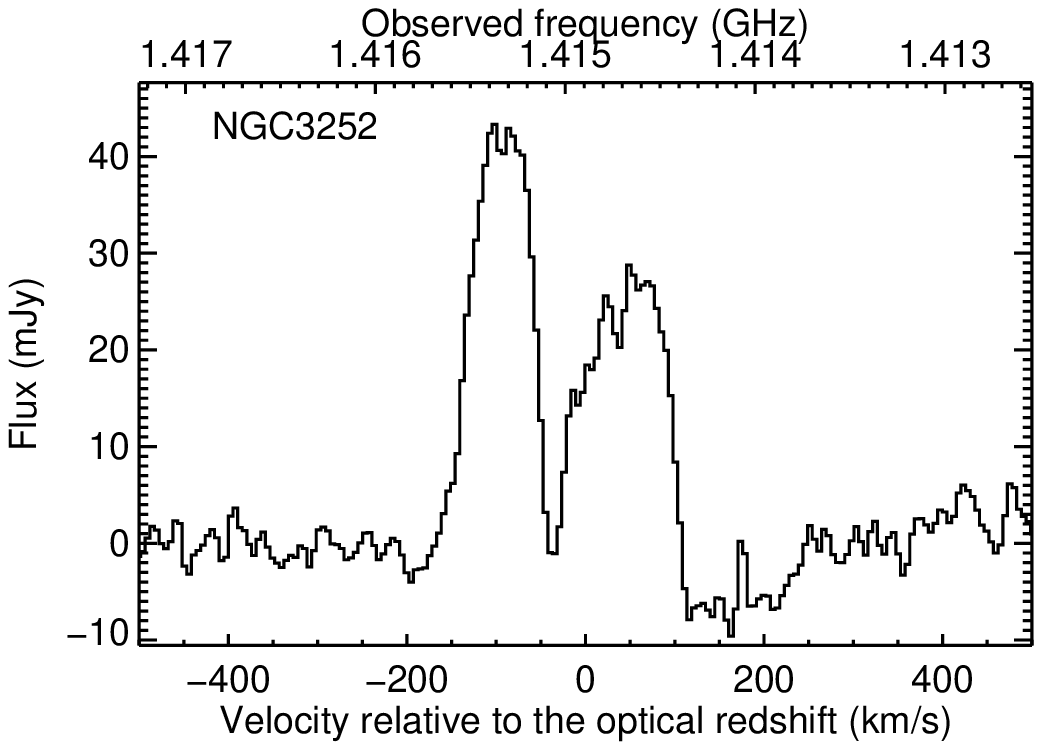} & 
\includegraphics[width=0.5\textwidth]{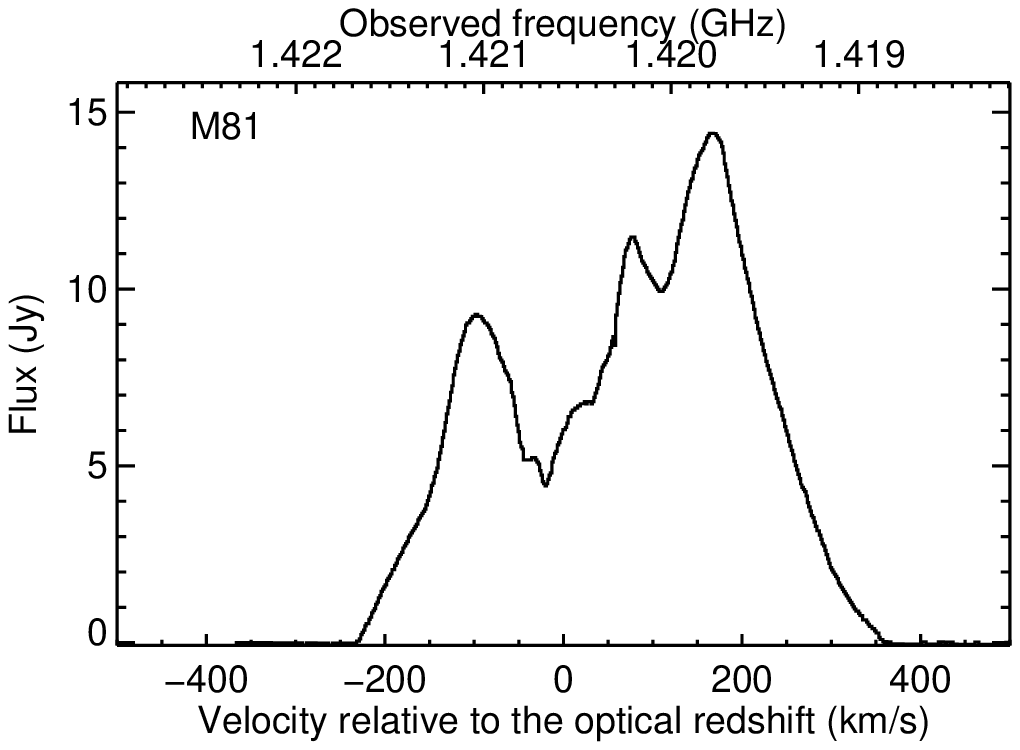}\\
\end{tabular}
\caption{{\hi} line spectra of {\ngc} and M81, the host of {\frb} and {\frbm} from \citet{masters14} and \citet{rots80}, respectively. Both spectra are highly asymmetric.
}
\label{fig:spec}
\end{figure*}

\begin{figure*}
\begin{tabular}{cc}
\includegraphics[width=0.5\textwidth]{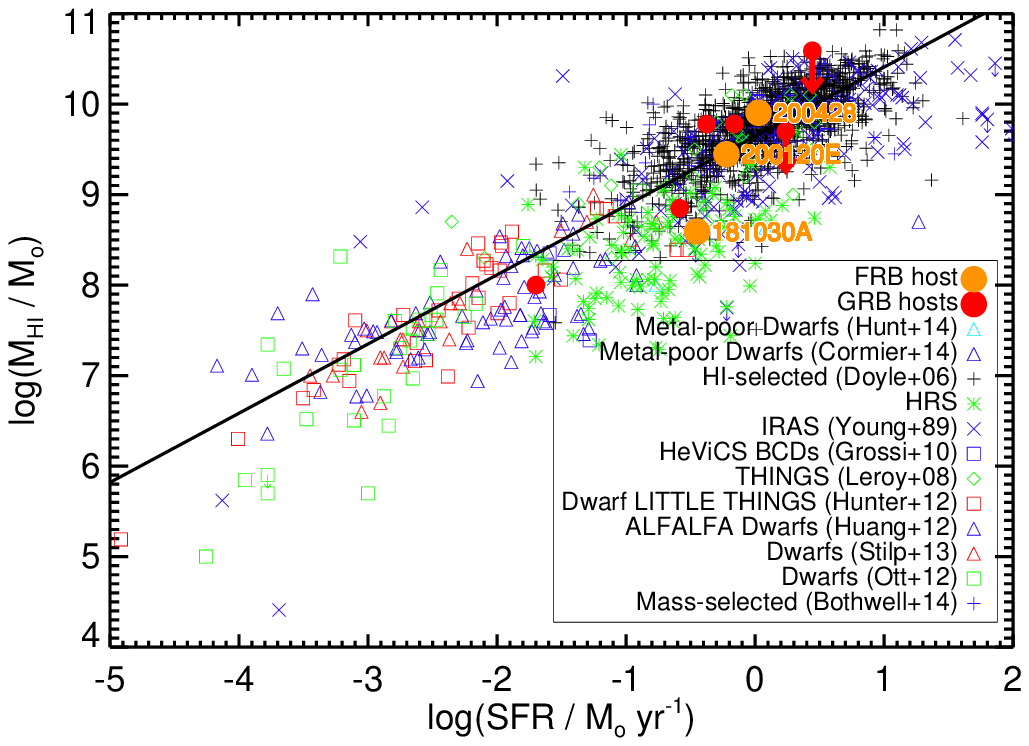} &
\includegraphics[width=0.5\textwidth]{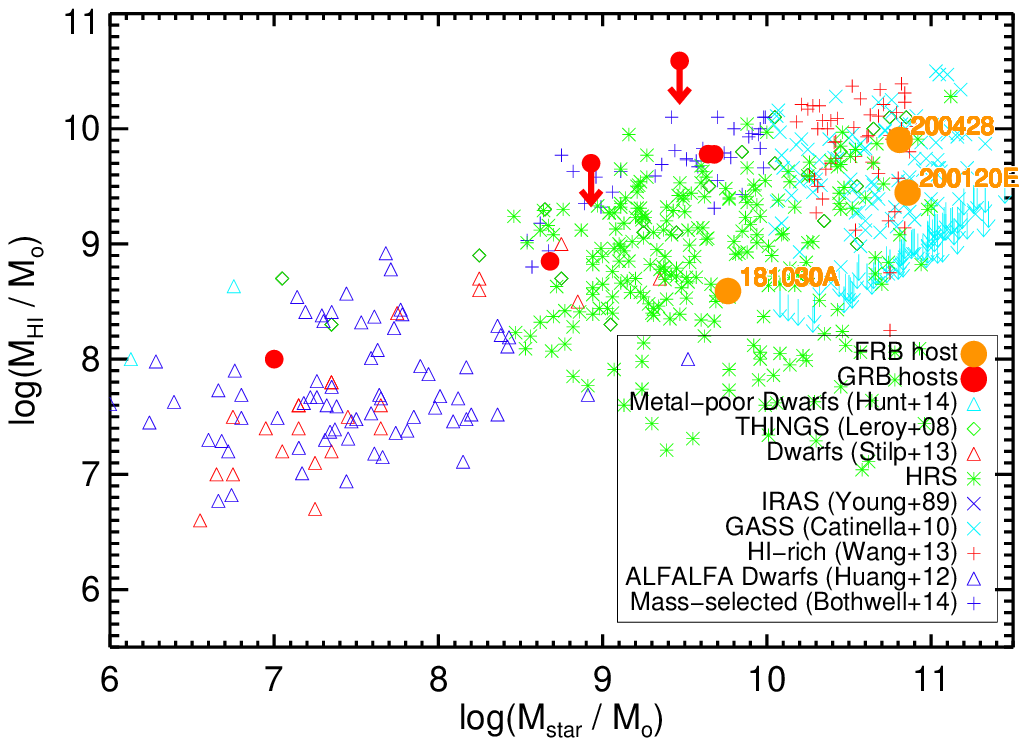} \\
\end{tabular}
\caption{Atomic gas mass as a function of SFR and stellar mass of hosts of FRBs (large orange circles), GRBs (red circles), and other star-forming galaxies (other symbols, as noted in the legend). The host of {\frb} has low atomic gas content for its SFR, but the hosts of the other two FRBs have normal atomic gas content. The figure is adopted from \citet{michalowski15hi}. 
}
\label{fig:mhi}
\end{figure*}

I use a cosmological model with $H_0=70$ km s$^{-1}$ Mpc$^{-1}$,  $\Omega_\Lambda=0.7$, and $\Omega_{\rm m}=0.3$. 
I use the FRB naming convention as for GRBs: YYMMDD with an additional letter indicating multiple FRBs discovered on the same day.

\begin{figure*}
\includegraphics[width=0.5\textwidth]{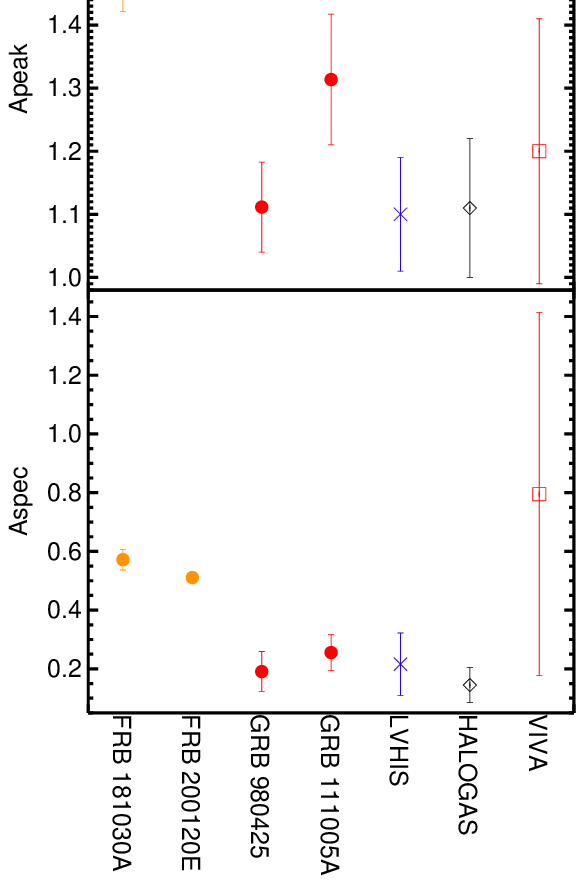}  
\caption{{\hi} line asymmetry diagnostics: the integrated flux ratio between the left and right halves of the spectrum, $A_{\rm flux}$ (eq.~6 in \citealt{reynolds20}); the flux ratio between  the left and right peaks of the spectrum, $A_{\rm peak}$ (eq.~7); and the residual from the subtraction  of the spectrum flipped around the flux-weighted velocity from the original spectrum, $A_{\rm spec}$ (eq.~8). The measurements for FRB and GRB hosts with {\hi} line detections are reported here for the first time, apart from GRB\,111005A \citep{lesniewska21}, and those for the galaxy surveys LVHIS and HALOGAS in low-density environments and VIVA in cluster high-density environments are from \citet{reynolds20}. The asymmetries of the FRB hosts are high, comparable to the highest asymmetries of galaxies in high-density environments.
}
\label{fig:asym}
\end{figure*}

\section{Sample and Data}
\label{sec:data}

I analyzed all FRB and GRB hosts which have {\hi} 21\,cm line detections. I searched for {\hi} data for all hosts of FRBs compiled  from the literature and the FRB Host Database\footnote{http://frbhosts.org} \citep{heintz20}. These are \object[NGC 3252]{\ngc} (UGC\,5732; PGC\,31278 2MASX\,J10342310+7345539), the host of \object[FRB 181030A]{\frb} \citep{bhardwaj21b}, \object{M81} (NGC\,3031), the host of \object[FRB 200120E]{\frbm} \citep{bhardwaj21,kirsten21,nimmo21,majid21}, and the \object{Milky Way}, the host of \object[FRB 200428]{\frbmw} \citep{bochenek20}.
They are the most local FRBs discovered to date and they are all repeating.

\citet{masters14} reported the {\hi} line flux of $5.89\pm0.14$\,Jy\,{\kms} for {\ngc}. Using eq.~2 in \citet{devereux90}, this corresponds to the atomic gas mass $\mhi=(3.90\pm0.09)\times10^8\,\msun$.
I used the star formation rate (SFR) and stellar mass (\mstar) of this galaxy from \citet{bhardwaj21b}.

M81, the host of {\frbm}, has an atomic gas mass $\mhi=(2.8\pm0.5)\times10^9\,\msun$ \citep{deblok18}, SFR of $(0.6\pm0.2)\,\msunyr$ \citep{gordon04}, and stellar mass of $\mstar=(7.2\pm1.7)\times10^{10}\,\msun$ \citep{deblok08,bhardwaj21}.		 For the Milky Way, the host of {\frbmw}, I used the following parameters: $\mhi=8\times10^9\,\msun$ \citep{kalberla09}, $\mbox{SFR}=(1.1\pm0.4)\msunyr$ \citep{robitaille10}, and $\mstar=(6.43\pm0.63)\times10^{10}\,\msun$ \citep{mcmillan11}.

I used the {\hi} spectrum of {\ngc}, the host of {\frb}, with the Green Bank Telescope (GBT) from \citet{masters14}. The beam has the size of $9'$, so the galaxy is unresolved. The spectral resolution is 10\,\kms.

For M81, the host of {\frbm}, I used the {\hi} spectrum from \citet{rots80}, also with GBT. Most of the optical disk of the galaxy is included within the beam. The spectral resolution is 22\,\kms.

The {\hi} lines of the hosts of both FRBs are shown on Fig.~\ref{fig:spec}. The total spectrum of the Milky Way is not used because it is difficult to reconstruct, given the uncertain distances of the detected {\hi} clouds.

The hosts of \object[GRB 980425]{GRB\,980425}, \object[GRB 060505]{060505}, and \object[GRB 111005A]{111005A} have been detected with {\hi} \citep{michalowski15hi,michalowski18grb,arabsalmani15b,arabsalmani19,lesniewska21}.
I used the total {\hi} spectra and integrated galaxy properties 
from \citet{michalowski15hi}. The {\hi} line asymmetry parameters for the latter are reported in \citet{lesniewska21}. 
This is not a homogeneous or representative sample, as it includes one GRB associated with a supernova and two supernova-less GRBs. They are all among the closest GRBs. Even for GRB\,060505 400\,Mpc away we do not expect any evolution of general galaxy properties, compared to local FRB hosts.

I compare the FRB hosts to other galaxy samples, as compiled in \citet{michalowski15hi}\footnote{\citet{young89b}, \citet{doyle06}, \citet{leroy08}, \citet{boselli10,boselli14}, \citet{catinella10}, \citet{grossi10},  \citet{cortese12b,cortese14}, \citet{hunter12}, \citet{huang12}, \citet{ott12}, \citet{stilp13}, \citet{wang13}, \citet{ciesla14}, \citet{cormier14}, \citet{hunt14b},  \citet{bothwell14}.}. 
This compilation includes local spirals, dwarfs, and (ultra)luminous infrared galaxies, as well as samples selected with {\hi} or stellar mass.

\section{Results}
\label{sec:results}

The integrated proprieties of {\hi}-detected galaxies are shown in Fig.~\ref{fig:mhi} (adopted from \citealt{michalowski15hi}): {\mhi} as a function of SFR and stellar mass.
{\ngc} has an atomic gas mass lower by a factor of 5 than the expectation given its SFR. The scatter of this relation is a factor of 2.4, so {\ngc} can be considered deficient in atomic gas (it is among 5\% of galaxies with the lowest {\mhi}/SFR ratio). 

The hosts of two other local FRBs are much different. M81, the host of {\frbm} is exactly consistent with the SFR-{\mhi} relation, whereas the Milky Way, the host of {\frbmw} is a factor of 1.7 higher. 
M81 and the Milky Way also have an order of magnitude higher stellar masses than  {\ngc}. The {\mstar}-{\mhi} diagram has a large scatter around a factor of 10, but all three FRB hosts are located close to the middle of the distribution.
This is different for GRB hosts, which are located close to the upper envelope of the distribution.
Indeed the mean $\mhi/\mstar$ ratios of GRB and FRB hosts are significantly different: $1.367\pm0.064$ and $0.077\pm0.025$, respectively.

The {\hi} lines of {\ngc} and M81 are clearly asymmetric with one of the wings having much higher peak flux (Fig.~\ref{fig:spec}; the {\hi} spectrum of the Milky Way is impossible to obtain directly). I quantified the {\hi} line asymmetry using the diagnostics defined by \citet{reynolds20}: 
the integrated flux ratio between the left and right halves of the spectrum, $A_{\rm flux}$ (eq.~6 in \citealt{reynolds20}); the flux ratio between  the left and right peaks of the spectrum, $A_{\rm peak}$ (eq.~7); and the residual from the subtraction  of the spectrum flipped around the flux-weighted velocity from the original spectrum, $A_{\rm spec}$ (eq.~8). 
I calculated the uncertainties of these parameters with a Monte Carlo simulation. I measured the standard deviation in line-free channels and perturbed each channel in the spectra 1000 times by the value drawn from a Gaussian distribution with such a width. I adopted the standard deviation of the asymmetry parameters in such synthetic spectra as their uncertainties.

I compared these values with measurements for galaxies with stellar masses of $9<\log(M_{\rm star}/M_\odot)<10$ from the Local Volume {\hi} Survey (LVHIS; \citealt{koribalski18}), the Hydrogen Accretion in Local Galaxies Survey (HALOGAS; \citealt{heald11}), and the VLA Imaging of Virgo in Atomic Gas (VIVA; \citealt{chung09}); see Fig.~8 and Table~3 of \citealt{reynolds20}. LVHIS and HALOGAS galaxies are in low environmental densities and VIVA galaxies are in high environmental densities (the Virgo cluster). 

The measurements for FRB and GRB hosts are reported here for the first time (Table~\ref{tab:asym}), apart from GRB\,111005A \citep{lesniewska21}.
The asymmetry parameters for the GRB\,060505 host cannot be measured because its spectrum has only two channels above the noise \citep{michalowski15hi}.
As shown in Fig.~\ref{fig:asym},
all asymmetry diagnostics for FRB hosts are 2--9 standard deviations higher than the mean for the low-density LVHIS and HALOGAS galaxies and even higher than the mean for the high-density VIVA galaxies, though consistent within errorbars with the latter. Only a few out of 150 galaxies in the sample of \citet{reynolds20} have such high asymmetry measurements.

In order to assess the deviation of FRB hosts from galaxies in low-density galaxies, I generated $10^9$ pairs of galaxies (reflecting two FRB hosts) from the asymmetry parameter distribution of HALOGAS galaxies, because this survey has smaller errorbars due to deeper data. None of these pairs turn out to have $A_{\rm flux}$ or $A_{\rm spec}$ equal or higher than those of the FRB hosts, implying probabilities of less than $10^{-9}$. Only 30 pairs have higher $A_{\rm peak}$ than FRB hosts, implying the probability of $3\times10^{-8}$. This implies that finding by chance even two galaxies with such asymmetric lines is almost impossible among low-density galaxies.

GRB hosts have much more symmetric {\hi} lines. The host of GRB\,111005A \citep{lesniewska21} has asymmetry diagnostics higher than those of low-density environments, but only by 1--2 standard deviations.
The GRB\,980425 host has a symmetric spectrum. The comparison of FRB and GRB hosts is, however, hampered by very small sample sizes of both populations.

In order to analyze the reason of these asymmetries, I have searched the  NASA/IPAC Extragalactic Database (NED) for companions within $60'$ (300\,kpc) and 1000\,\kms.
{\ngc} is a member of a galaxy group \citep[number 112989;][]{kourkchi17,shaya17}, as also noted by \citet{bhardwaj21b}. It has a dynamical mass of $2\times10^{11}\,\msun$ and a radius of 0.135\,Mpc. 
M81 is a member of a well known interacting group \citep[e.g.][]{deblok18}. Finally, the Milky Way is interacting with the Magellanic Clouds and other dwarf galaxies, whereas the Andromeda galaxy, with a similar mass, is only 750\,kpc away.

\begin{table*}
\centering
\caption{{\hi} line asymmetry parameters as defined in \citet{reynolds20}.}
\begin{tabular}{lccc}
\hline\hline
Galaxy & $A_{\rm flux}$ & $A_{\rm peak}$ & $A_{\rm spec}$\\
\hline
FRB 181030A & $1.314\pm0.072$ & $1.505\pm0.084$ & $0.5719\pm0.0351$ \\
FRB 200120E & $1.505\pm0.002$ & $1.551\pm0.004$ & $0.5108\pm0.0005$ \\
\hline
GRB 980425 & $1.027\pm0.066$ & $1.112\pm0.071$ & $0.1907\pm0.0684$ \\
GRB 111005A & $1.237\pm0.102$ & $1.314\pm0.104$ & $0.2555\pm0.0610$ \\
\hline
LVHIS & $1.110\pm0.070$ & $1.100\pm0.090$ & $0.2160\pm0.1060$ \\
HALOGAS & $1.070\pm0.050$ & $1.110\pm0.110$ & $0.1450\pm0.0590$ \\
VIVA & $1.190\pm0.170$ & $1.200\pm0.210$ & $0.7950\pm0.6180$ \\
\hline
\end{tabular}
\label{tab:asym}
\end{table*}

\section{Discussion}
\label{sec:discussion}

To summarize, I report three observables: {\it i)} all three FRB hosts are interacting galaxies; {\it ii)} the {\hi} spectra of both FRB hosts with such data available are highly asymmetric, several standard deviations above the general population of galaxies outside galaxy clusters (Fig.~\ref{fig:asym}); {\it iii)} two FRB hosts have normal atomic gas properties and one is strongly deficient in atomic gas (Fig.~\ref{fig:mhi}). If this holds for larger samples of FRB hosts with {\hi} observations, then this will indicate that FRBs are connected with a recent enhancement of star formation due to interaction. Hence, this result supports the fast FRB channel mentioned above. One of the possibility would be a connection with massive stars that explode as type II SNe, with  delay times short enough that signatures of interaction giving rise to the birth of their progenitors should still be visible. Indeed, \citet{heintz20} found that the properties of FRB hosts are best matched to that of core-collapse SNe. This is also consistent with the clear association of an FRB with a Galactic magnetar \citep{bochenek20}.
Moreover, FRB hosts are inconsistent with being drawn from the
stellar mass distribution of star-forming galaxies \citep{heintz20}, which rules out some of the slow FRB channels with rates proportional to stellar mass.

On the other hand, there are several observations that do not support  the connection of FRB with recent enhancement of star formation. First, the location of {\frbm} in a globular cluster suggests a connection with an old stellar population  \citep{kirsten21}. Moreover, \citet{fong21} measured approximately constant SFR of the host of FRB\,201124A. However, this measurement is global for an entire galaxy, whereas the enhancement may be local. Indeed the persistent radio source consistent with that FRB location has been interpreted as the emission of a star forming region \citep{fong21,ravi21}.

It is intriguing that GRB hosts do not exhibit such line asymmetries. The current samples of hosts of FRBs and GRBs are too small to draw firm conclusions, but if this holds for larger samples, then this will advocate different progenitors. It has been claimed that a large fraction of the atomic gas of GRB hosts has recently arrived from the intergalactic medium (IGM), which caused an enhancement of star formation and lowered metallicity, allowing the GRB progenitors to form \citep{michalowski15hi,michalowski16,michalowski18co}. The difference in {\hi} asymmetry between GRB and FRB hosts can be explained if this gas inflow for GRB hosts is more gentle, not disturbing the velocity field of galaxies. Indeed, M81 is strongly interacting with M82, which is only four times less massive, whereas GRB hosts are usually not found in interacting groups.
This scenario is consistent with low metallicity being a requirement for the birth of a GRB progenitor, because gas in interacting groups cannot be expected to have low metallicity, whereas this is the case for gas inflow from the IGM for isolated galaxies. Indeed, FRBs have been found not to avoid high metallicity environments \citep{heintz20}. 

Another explanation of the difference is connected with stellar masses. GRB hosts are much less massive than FRB hosts (see Fig.~\ref{fig:mhi} here and Fig.~10 in \citealt{heintz20}). Such less massive galaxies have a smaller degree of ordered rotational motion of gas, so their {\hi} lines are generally consistent with single Gaussians, and as such, are more symmetric. This can only be investigated with {\hi} observations of FRB and GRB hosts matched in stellar mass.
In either case, the difference in {\hi} properties of FRB and GRB hosts suggests different mechanisms for these two classes of transients.

{\hi} line observations are needed to investigate the interaction with other galaxies, because the asymmetric line profiles and/or the existence of atomic gas tidal tails or bridges are definitive evidence of interaction \citep{deblok18,michalowski20b,reynolds20,watts20}.
Interactions are more robustly identified using {\hi} data also because if the merging companion is a dwarf galaxy, then it may be too faint in the optical.

The group containing {\ngc} is relatively small, but the interaction with other members may be able to remove the atomic gas from {\ngc}, explaining its factor of 5 deficiency in atomic gas. Indeed, it has been found that galaxies with asymmetric {\hi} lines exhibit lower atomic gas content due to interactions \citep{watts20,reynolds20,hu21}.
With such a small sample, it is difficult to know whether this is usual for FRB hosts.

Apart from the sample size, this analysis has some caveats. First, different FRBs may have different emission mechanisms, which would blur the picture painted by their hosts when analyzed together. Indeed, {\frbmw} in the Milky Way was connected with a magnetar and was 30 times fainter than the faintest extragalactic FRB \citep{bochenek20}, whereas {\frbm} was localized in a globular cluster \citep{kirsten21}. On the other hand, all FRBs in this sample are repeaters, so no conclusion can be drawn from this about the nature of the nonrepeating population. Hence, with a larger sample such studies should be done with subcategories. Second, local FRBs (and their hosts) analyzed here, may be different from more distant ones. Indeed, the three FRB hosts analyzed here are some of the most massive FRB hosts \citep[see Fig.~7 of][]{heintz20}. This will be addressed with the Square Kilometre Array (SKA), able to detect {\hi} at higher redshifts. The galaxy merger rate was higher at high redshifts \citep{deravel09}, so if the FRB-interaction connection holds at high redshifts, then the FRB rate should increase with redshift. 
Finally, I analyzed here only the spatially integrated properties of FRB hosts. In fact, the properties of the local environment of an FRB progenitor should be more strongly connected with the emission mechanism.

Hence, this study can be improved with resolved {\hi} observations of FRBs with precise localizations.
It can also be extended to hosts of other transients. I note that the hosts of SNe type Ic-BL \object[SN 2002ap]{2002ap} \citep{kamphuis92,michalowski20} and \object[SN 2009bb]{2009bb} \citep{michalowski18} as well as of a fast blue optical transient \object[AT 2018cow]{AT\,2018cow} \citep{michalowski19,roychowdhury19} have very symmetric line profiles.

\section{Conclusions}
\label{sec:conclusion}

I present the first analysis of atomic gas properties of FRB hosts. I found very asymmetric {\hi} line profiles, compared to other star-forming galaxies, and atomic gas deficiency in one of the FRB hosts. All three FRB hosts are members of interacting groups. This suggests the connection between FRBs and the increased star formation activity due to interaction. Hence, this supports fast FRB channels in which the delay between the birth of the progenitor of the burst is of the order of tens or hundreds of million years, not billion of years. GRB hosts do not exhibit such strong {\hi} line asymmetries, so the formation of their progenitors does not seem to be connected with strong galaxy interactions. This also suggests that the emission mechanisms of FRBs and GRBs are likely different.

\acknowledgments 

I would like to thank 
the referee for useful suggestions, and 
Joanna Baradziej, 
for discussions and comments.
I acknowledge the support of 
the National Science Centre, Poland through the SONATA BIS grant 2018/30/E/ST9/00208.
This research has made use of 
the NASA/IPAC Extragalactic Database (NED) which is operated by the Jet Propulsion Laboratory, California Institute of Technology, under contract with the National Aeronautics and Space Administration;
Edward Wright cosmology calculator \citep{wrightcalc};
and NASA's Astrophysics Data System Bibliographic Services.



\end{document}